\documentclass[12pt]{article}

\usepackage{epsfig}

\usepackage{graphicx}

\def\square{\kern1pt\vbox{\hrule height 1.2pt
\hbox{\vrule width 1.2pt\hskip 3pt
\vbox{\vskip 6pt}\hskip 3pt\vrule width 0.6pt}
\hrule height 0.6pt}\kern1pt}
\def\ltwid{\mathrel{\raise.3ex\hbox{$<$\kern-.75em\lower1ex\hbox{$\sim$}}}}

\begin{document}

\begin{titlepage}
\begin{flushright}
CRETE-09-13 \\ UFIFT-QG-09-03
\end{flushright}

\vspace{0.5cm}

\begin{center}
\bf{Post-Inflationary Evolution via Gravitation}
\end{center}

\vspace{0.3cm}

\begin{center}
N. C. Tsamis$^{\dagger}$
\end{center}
\begin{center}
\it{Department of Physics, University of Crete \\
GR-710 03 Heraklion, HELLAS.}
\end{center}

\vspace{0.2cm}

\begin{center}
R. P. Woodard$^{\ast}$
\end{center}
\begin{center}
\it{Department of Physics, University of Florida \\
Gainesville, FL 32611, UNITED STATES.}
\end{center}

\vspace{0.3cm}

\begin{center}
ABSTRACT
\end{center}
\hspace{0.3cm} We study a class of non-local, purely
gravitational models which have the correct structure
to reproduce the leading infrared logarithms of quantum
gravitational back-reaction during the inflationary
regime. These models end inflation in a distinctive phase 
of oscillations with slight and short violations of the 
weak energy condition and should, when coupled to matter, 
lead to rapid reheating. By elaborating this class of 
models we exhibit one that has the same behaviour during 
inflation, goes quiescent until the onset of matter 
domination, and induces a small, positive cosmological 
constant of about the right size thereafter.

\vspace{0.3cm}

\begin{flushleft}
\end{flushleft}

\vspace{0.1cm}

\begin{flushleft}
$^{\dagger}$ e-mail: tsamis@physics.uoc.gr \\
$^{\ast}$ e-mail: woodard@phys.ufl.edu
\end{flushleft}

\end{titlepage}

${\bullet \; \;}$ {\bf Introduction:} 
During the inflationary era infrared gravitons are 
produced out of the vacuum because of the accelerated
expansion of spacetime. The interaction stress among
the gravitons produced -- an inherently non-local effect 
-- can lead to a non-trivial quantum gravitational 
back-reaction on inflation \cite{NctRpw1}. In a previous
paper \cite{NctRpw2} we proposed a phenomenological
model which can provide evolution beyond perturbation 
theory. In one sentence, we constructed an {\it effective}
conserved stress-energy tensor $T_{\mu\nu}[g]$ which
modifies the gravitational equations of motion:
\footnote{Hellenic indices take on spacetime values while 
Latin indices take on space values. Our metric tensor 
$g_{\mu\nu}$ has spacelike signature and our curvature 
tensor equals:
$R^{\alpha}_{~\beta\mu\nu} \equiv 
\Gamma^{\alpha}_{~\nu\beta, \mu} +
\Gamma^{\alpha}_{~\mu\rho} \;
\Gamma^{\rho}_{~\nu\beta} -
(\mu \leftrightarrow \nu)$. 
The initial Hubble constant is $3H^2_0 \equiv \Lambda$. 
We restrict our analysis to scales 
$M \equiv (\, \Lambda / 8 \pi G \,)^{\frac14}$ 
below the Planck mass $M_{\rm Pl} \equiv G^{-\frac12}$ 
so that the dimensionless coupling constant 
$\epsilon \equiv G \Lambda$ of the theory is 
small.}
\begin{equation}
G_{\mu\nu} \; \equiv \;
R_{\mu\nu} \, - \, \frac12 g_{\mu\nu} \, R \; = \;
- \Lambda \, g_{\mu\nu} \, + \,
8 \pi G \, T_{\mu\nu}[g] 
\;\; . \label{eom1}
\end{equation}
and which, we hope, contains the most cosmologically
significant part of the full effective quantum
gravitational equations.

Our physical {\it ansatz} consisted of parametrizing
$T_{\mu\nu}[g]$ as a ``perfect fluid'':
\begin{equation}
T_{\mu\nu}[g] \; = \;
(\rho + p) \, u_{\mu} \, u_{\nu} \, + \,
p \, g_{\mu\nu}
\;\; , \label{Tmn}
\end{equation}
with the gravitationally induced pressure given
as the following functional of the metric tensor:
\begin{equation}
p[g](x) \; = \;
\Lambda^2 \, f[- \epsilon \, X](x) 
\qquad , \qquad
X \, \equiv \, \frac{1}{\square} \, R
\;\; , \label{pressure}
\end{equation}
where the function $f$ grows without bound and 
satisfies:
\begin{equation}
f[- \epsilon \, X] \; = \;
- \epsilon \, X \, + \, O(\epsilon^2)
\;\; , \label{fincr}
\end{equation}
and where the scalar d'Alembertian:
\begin{equation}
\square \, \equiv \;
\frac{1}{\sqrt{-g}} \;
\partial_{\mu} \Big( \,
g^{\mu\nu} \sqrt{-g} \; \partial_{\nu} \, \Big)
\;\; , \label{box}
\end{equation}
is defined with retarded boundary conditions.
The induced energy density $\rho[g]$ and 4-velocity
$u_{\mu}[g]$ were determined, up to their initial
value data, from stress-energy conservation:
\begin{equation}
D^{\mu} \, T_{\mu\nu} \; = \; 0
\;\; . \label{cons1}
\end{equation}
The 4-velocity was chosen to be timelike and
normalized:
\begin{equation}
g^{\mu\nu} \, u_{\mu} u_{\nu} = -1
\qquad \Longrightarrow \qquad
u^{\mu} \, u_{\mu ; \nu} = 0
\;\; . \label{u}
\end{equation}
We cannot exhibit $u_{\mu}[g](x)$ and $\rho[g](x)$ 
as explicit functionals of the metric the way 
expression (\ref{pressure}) does for the pressure, 
but it is straightforward to solve the conservation 
equations for a homogeneous and isotropic background
\footnote{The line element in co-moving coordinates is 
$ds^2 = -dt^2 + a^2(t) \, d{\vec x} \cdot d{\vec x}$.
In terms of the scale factor $a$, the Hubble parameter 
equals $H(t) = {\dot a} \, a^{-1}$ and the deceleration 
parameter equals $q(t) = - a \, {\ddot a} \,
{\dot a}^{-2}$.}
and even for perturbations \cite{NctRpw3}. For the
analysis that will follow we require only the background 
results:
\begin{equation}
u_{\mu}(t) \; = \; -a(t) \, \delta^{~0}_{\mu} 
\qquad , \qquad
\rho(t) \; = \; -p(t) \, + \,
\int_0^t dt' \; \dot{p}(t') 
\left[ \frac{a(t')}{a(t)} \right]^3 
\;\; . \label{rho}
\end{equation}
Hence the equation of state of our ansatz for the induced 
stress-energy is:
\begin{equation}
w(t) \; \equiv \; 
\frac{p(t)}{\rho(t)} \; = \;
- \left[ \, 1 \, - \,
\int_0^t dt' \; \frac{\dot{p}(t')}{p(t)} \;
\frac{a^3(t')}{a^3(t)} \, \right]^{-1}
\;\; . \label{w}
\end{equation}

The homogeneous and isotropic evolution
of this model -- using a combination of numerical 
and analytical methods -- revealed the following 
basic features
\footnote{In \cite{NctRpw2}, our analytical results 
were obtained for any function $f$ satisfying 
(\ref{fincr}) and growing without bound, our numerical 
results for the choice: $f(x) = \exp(x) - 1$.} 
: \\
{\it 01.} After the onset and during the era of
inflation, the source $X(t)$ grows while the curvature
scalar $R(t)$ and Hubble parameter $H(t)$ decrease.

\vspace{0.1cm}

\noindent
{\it 02.} Inflationary evolution dominates roughly 
until we reach a critical point $X_{cr}$ defined by:
\begin{equation}
1 - 8 \pi G \Lambda \, f[ - G \Lambda \, X_{cr}] 
\; \equiv \; 0
\;\; . \label{Xcr}
\end{equation}
{\it 03.} The epoch of inflation ends close to but 
before the universe evolves to the critical time.
This is most directly seen from the deceleration
parameter since initially $q(t=0) = -1$ while at 
criticality $q(t=t_{cr}) = +\frac12$.

\vspace{0.1cm}

\noindent
{\it 04.} Oscillations in $R(t)$ become significant 
as we approach the end of inflation; they are centered 
around $R = 0$, their frequency equals:
\begin{equation}
\omega \; = \; 
G \Lambda H_0 \sqrt{72 \pi \, f_{cr}'}
\;\; , \label{omega}
\end{equation}
and their envelope is linearly falling with time.

\vspace{0.1cm}

\noindent
{\it 05.} During the oscillations era, although there
is net expansion, the oscillations of $H(t)$ take it
to small negative values for small time intervals --
a feature conducive to rapid reheating; those of 
${\dot H}(t)$ take it to positive values for about 
half the time; and, those of $a(t)$ are centered 
around a linear increase with time. 

\vspace{0.1cm}

Since we shall be concerned with the various 
post-inflationary phases to which the universe
evolved, and since all these phases are characterized
by constant $\varepsilon \equiv -{\dot H} H^{-2} =
1 + q$ , it is convenient to provide expressions 
for some basic quantities describing such spacetimes.
In particular, in terms of their initial values, the 
time evolution of the scale factor, Hubble parameter 
and Ricci scalar are, respectively:
\begin{eqnarray}
\varepsilon \; \equiv \;
- \frac{{\dot H}(t)}{H^2(t)}  
\quad \Longrightarrow \qquad
a(t) \!\!& = &\!\!
a_{in} \,
\Big[ \, 1 + \varepsilon \, H_{in}
(t - t_{in}) \, \Big]^{\frac{1}{\varepsilon}}
\; \; , \label{aeps} \\
H(t) \!\!& = &\!\!
\frac{H_{in}}
{1 + \varepsilon \, H_{in}
(t - t_{in})}
\; \; , \label{Heps} \\
R(t) \!\!& = &\!\! 
6 {\dot H} + 12 H^2
\; = \;
6 \, (2 - \varepsilon) \, H^2(t)
\; \; . \qquad \label{Reps} 
\end{eqnarray}
We shall also need the temporal component of the 
Ricci tensor:
\begin{equation}
R_{00} \; = \;
+3 \, q \, H^2 
\; = \;
-3 \, (1 - \varepsilon) \, H^2
\;\; . \label{R00eps}
\end{equation}
Moreover, the first two time derivatives of the 
curvature scalar are:
\begin{eqnarray}
{\dot R} \!\!& = &\!\!
-12 \varepsilon \, (2 - \varepsilon) \, H^3
\;\; , \label{dotReps} \\
{\ddot R} \!\!& = &\!\!
+36 \varepsilon^2 \, (2 - \varepsilon) \, H^4
\;\; , \label{ddotReps}
\end{eqnarray}
so that:
\begin{equation}
\square R \; = \;
-{\ddot R} - 3 H {\dot R} \; = \;
+36 \varepsilon \, (1 - \varepsilon)
(2 - \varepsilon) \, H^4
\;\; . \label{boxReps}
\end{equation}
Notice that $R$ vanishes for radiation $(\varepsilon = 2)$
and $\square R$ for radiation as well as inflation
$(\varepsilon = 0)$. \\

${\bullet \; \;}$ {\bf Two Problems:} 
The homogeneous and isotropic evolution described by the 
simple model briefly reviewed in the introduction does 
not give a completely satisfactory end to inflation. The 
oscillations that occur after the end of inflation are 
not a problem, but the average expansion $a(t) \propto t$ 
is unacceptably rapid. At that rate there would be no 
reheating and the late time universe would be cold and 
empty. Nonetheless, the same is true for scalar-driven 
inflation if one ignores the possibility for energy to 
flow from the inflaton into ordinary matter. We believe 
that energy will flow from the gravitational sector of our 
model into ordinary matter to create a radiation dominated 
universe, just as it is thought to do for scalar-driven 
inflation. 

An amazing possibility arises if this process can be shown 
to occur: {\it our quantum gravitational correction cancels 
the bare cosmological constant and then becomes dormant 
during the epoch of radiation domination.} To see this, 
suppose the deceleration parameter has the pure radiation 
value of $q(t) = +1$ for times $t > t_r$.  This case
corresponds to $\varepsilon = 2$ and (\ref{aeps}-\ref{Reps})
give:
\begin{eqnarray}
q = +1 
\quad \Longrightarrow \qquad 
a(t) \!\!& = &\!\!
a_r \, \Bigl[ \,
1 + 2 H_0 (t - t_r)
\, \Bigr]^{\frac12} 
\;\; , \hspace{3cm} \label{arad} \\
H(t) \!\!& = &\!\!
\frac{H_r}{1 + 2 H_r (t - t_r)} 
\;\; , \label{Hrad} \\
R(t) \!\!& = &\!\!
0
\;\; , \label{Rrad}
\end{eqnarray}
Our simple source $X(t)$ obeys the differential equation 
$\square X = R$, so for $t > t_r$ it must be a linear 
combination of its two homogeneous solutions:
\begin{eqnarray}
\forall t \!\!& > &\!\!t_r 
\quad \Longrightarrow \qquad 
\square X \; = \; 0
\quad \Longrightarrow \qquad 
\nonumber \\
X(t) \!\!& = &\!\!
X_r \, + \,
\dot{X}_r \int^t dt'
\left[ \frac{a_r}{a(t')} \right]^3 
\, = \; 
X_r \; - \; \frac{\dot{X}_r}{H_r} \;
\frac{1}{\sqrt{1 + 2 H_r (t - t_r)}} 
\;\; . \qquad \label{Xrad}
\end{eqnarray}
The only solution consistent with $q = +1$ is:
\footnote{For all power laws, such as radiation, the 
pressure and density fall like $t^{-2}$. Therefore, 
the second of the two homogeneous solutions cannot 
be present, implying that ${\dot X}_r = 0$, since 
its time dependence is $t^{-\frac12}$ and could not 
sustain radiation. Neither can the first homogeneous 
solution sustain radiation unless the constant $X_r$ 
eliminates the cosmological constant $3 H_0^2$ and, 
hence, equals $X_{cr}$.}
\begin{equation}
X_{r} \; = \; X_{cr} 
\qquad , \qquad
\dot{X}_r \; = \; 0
\;\; . \label{Xradsol} 
\end{equation}
Note from equation (\ref{pressure}) that $X(t)$ 
becoming constant implies that the induced pressure 
$p(t)$ is also constant. One can see from equations 
(\ref{rho}-\ref{w}) that the energy density also 
becomes constant provided the approach to radiation 
domination results in the following condition:
\begin{equation}
\int_0^{t_r} dt \; \dot{p}(t) \, a^3(t) 
\; = \; 0 
\;\; . \label{fixed}
\end{equation}
In fact this is also the condition for achieving the 
fixed point solution (\ref{Xradsol}), and it results 
in our induced stress-energy becoming perfect vacuum 
energy which completely cancels the bare cosmological 
constant $\Lambda$.

Having $X(t)$ approach $X_{cr}$ within the context 
of a hot, radiation dominated universe would be a 
great success for our model, but the eventual 
transition to matter domination poses enormous 
problems. The onset of matter domination is really 
a gradual process but let us simplify the exposition 
by considering a sudden change from $q = +1$ to 
$q = + \frac12$ at some time $t_m \gg t_r$. During 
this matter dominated epoch, for which $\varepsilon
= \frac32$, expressions (\ref{aeps}-\ref{Reps})
become:
\begin{eqnarray}
q = +\frac12 
\quad \Longrightarrow \qquad 
a(t) \!\!& = &\!\!
a_m \, \Bigl[ \,
1 + \frac32 H_m (t - t_m) \, \Bigr]^{\frac23} 
\;\; , \hspace{3cm} \label{amat} \\
H(t) \!\!& = &\!\!
\frac{H_m}{1 + \frac32 H_m (t - t_m)} 
\;\; , \label{Hmat} \\
R(t) \!\!& = &\!\!
\frac{3 H_m^2}{\Big[ \,
1 + \frac32 H_m (t - t_m) \, \Big]^2} 
\;\; . \label{Rmat}
\end{eqnarray}
where $H_m$ and $a_m$ are $H(t_m)$ and $a(t_m)$, 
respectively, computed from the radiation dominated 
geometry (\ref{arad}-\ref{Hrad}). The resulting 
change in the source $X(t)$ is:
\begin{eqnarray}
q = +\frac12 
\;\; & \Longrightarrow & 
\nonumber \\
\Delta X(t) \!\!& \equiv &\!\!
X(t) - X_{cr} 
\; = \;
- \frac43 \ln \Bigl[ \,
1 + \frac32 H_m (t - t_m) \, \Bigr] 
\; + \; O(1) 
\;\; . \qquad \label{Xmat}
\end{eqnarray}
The fact that matter domination causes $X(t)$ to 
move off its critical value $X_{cr}$ means that 
our induced stress-energy no longer behaves like 
perfect vacuum energy. That would pose no problem 
for cosmology provided that: \\
{\it (i)} the residual effect from the perturbation 
$\Delta X(t)$ results in a pressure and energy density 
which are initially small compared to the energy 
density of matter, \\
{\it (ii)} the residual effect eventually acts to 
induce a phase of acceleration after a suitably 
long time lag.

To understand what is wrong with the change (\ref{Xmat}) 
caused by matter domination, it is useful to recall our 
{\it ansatz} for the quantum gravitationally induced 
pressure:
\begin{equation}
p[g](x) \; = \;
\Lambda^2 \, f [-G\Lambda \, X](x) 
\;\; . \label{p}
\end{equation}
In the context of this ansatz there are two major problems 
with (\ref{Xmat}):

\vspace{0.2cm}

{\it 01. The sign problem.}  It derives from the function 
$f(x)$ in (\ref{p}) being monotonically increasing and 
unbounded. Hence, pushing $X(t)$ below $X_{cr} \ll 0$ 
results in positive total pressure, whereas observation 
implies negative pressure during the current epoch 
\cite{riess, wang}. Note that we cannot alter this feature 
of $f(x)$ without sacrificing the very desirable ability 
of the model to cancel an arbitrary bare cosmological 
constant. 

\vspace{0.2cm}

{\it 02. The magnitude problem.}  In one sentence, the
magnitude of the total pressure produced by (\ref{Xmat})
is vastly too large. The problem arises from the factors 
of the bare cosmological constant $\Lambda$ in our ansatz 
(\ref{p}). The total pressure $p_{\rm tot}$ is the sum 
of the classical contribution and our ansatz (\ref{p}):
\begin{eqnarray}
p_{\rm tot} \!\!& = &\!\!
- \, \frac{\Lambda}{8\pi G} \, \Biggl\{
1 - 8 \pi G \Lambda \;
f\Big[ -G \Lambda \, (X_{cr} + \Delta X) \, \Big] \Biggr\} 
\label{ptot1} \\
\!\!& \simeq &\!\!
- \, \frac{\Lambda}{G}
\times (G \Lambda)^2 \, f_{cr}' \; \Delta X 
\;\; . \label{ptot2}
\end{eqnarray}
Comparing with the currently observed value $p_{\rm now}$ 
of the pressure:
\begin{equation}
p_{\rm now} \; \simeq \; 
- \, \frac{3}{8 \pi G} \, H_{\rm now}^2
\;\; , \label{pnow}
\end{equation}
gives:
\begin{equation}
\frac{p_{\rm tot}}{p_{\rm now}} 
\; \simeq \;
\left( \frac{G \Lambda \, H_0}{H_{\rm now}}
\right)^2 f_{cr}' \; \Delta X 
\; \simeq \;
10^{86} \times f_{cr}' \times \Delta X 
\; , \label{pratio}
\end{equation}
where we have assumed $H_0 \sim 10^{13}~{\rm GeV}$ 
and $H_{\rm now} \sim 10^{-33}~{\rm eV}$. The derivative 
$f_{cr}'$ is unity for the linear model and of order 
$(G \Lambda)^{-1} \sim 10^{12}$ for the exponential 
model, so we expect $f_{cr}'$ to be at least of order 
one and possibly much greater. 

\vspace{0.2cm}

There is no way of addressing either problem without 
generalizing our ansatz (\ref{p}) for the pressure. 
This necessarily takes us away from what can be 
motivated by explicit computation during the de Sitter 
regime. \\

${\bullet \; \;}$ {\bf Decreasing the Magnitude:} 
The magnitude problem arises because the constant $\Lambda$ 
in (\ref{p}) is about the square of the inflationary Hubble 
parameter rather than its late time descendant that could 
be $55$ orders of magnitude smaller. Solving the problem 
entails replacing one of these factors of $\Lambda$ by some 
dynamical scalar that changes as time evolves in a way that
also preserves the original relaxation mechanism. The latter
requirement rules out any tampering with the factor of 
$\Lambda^2$ that multiplies $f[-G \Lambda \, X]$ in (\ref{p}). 
It is safer to make the factor of $\Lambda$ in the argument 
of the function $f$ dynamical and move it to the right of 
the $\square^{-1}$:
\begin{equation}
-G \Lambda \, X \; = \;
- \frac{G \Lambda}{\square} \, R  
\qquad \longrightarrow \qquad
- \frac{G \Lambda}{\square}
\left( R \times \frac{S}{\Lambda} \, \right) 
\; \equiv \;
-G \Lambda \times Y[g] 
\;\; ,\label{changeX}
\end{equation}
where $S$ is an appropriate scalar quantity. The idea is 
for (\ref{changeX}) to approach $ -G \Lambda \times X_{cr}$ 
during inflation and then freeze in to this value during 
radiation domination, throughout which the scalar falls 
off so that subsequent evolution is driven by an acceptably 
small source. There are many possibilities for the scalar 
$S$ in (\ref{changeX}), which we shall always normalize 
so as to make the ratio $S\Lambda^{-1}$ give unity for 
de Sitter spacetime. 

One might think $S = \frac14 R$ can work, but numerical 
simulations show that it responds too quickly to the 
slowing geometry. Instead of inflation ending, the scale 
factor approaches an accelerating, power law expansion 
-- that is, $a(t) \propto t^s$ with $s > 1$. It is 
instructive to understand why this happens by perturbing 
the pressure around the critical value which cancels 
the bare cosmological constant. The relevant $FRW$ 
equation of motion becomes:
\begin{eqnarray}
-2\dot{H} - 3 H^2 \!\!& = &\!\!
-\Lambda + 8 \pi G \, p
\label{gijeqn} \\
& = &\!\!
-\Lambda \, \Biggl\{ \,
1 - 8 \pi G \Lambda \,
f\Big[ -G\Lambda ( X_{cr} + \Delta Y ) \, \Big] \, \Biggr\} 
\label{gijeqn2} \\
& \approx &\!\!
-8 \pi (G \Lambda)^2 \, f_{cr}' \; \Lambda 
\times \Delta Y 
\;\; .\label{gijeqn3}
\end{eqnarray}
When $S = \frac14 R$, the function $\Delta Y$ obeys:
\begin{equation}
S = \frac14 \, R 
\quad \Longrightarrow \qquad
\square \Delta Y \; = \; \frac{R^2}{4 \Lambda} 
\;\; . \label{deltaY}
\end{equation}
Then, for $a(t) \propto t^s$ we have:
\begin{eqnarray}
a(t) \, \propto \, t^s 
& \Longrightarrow &\quad 
R \; = \; \frac{6s \, (2s-1)}{t^2}
\nonumber \\
& \Longrightarrow &\quad 
\Delta Y \; = \;
\frac{Y_1}{t^{3s-1}} \, + \,
\frac{3 s^2 (2s - 1)^2}{2(s - 1) \Lambda t^2} 
\;\; , \label{deltaY2}
\end{eqnarray}
where $Y_1$ is an integration constant. For $s > 1$ the 
homogeneous solution proportional to $Y_1$ falls off 
faster than the inhomogeneous solution. Neglecting the 
homogeneous solution and substituting into (\ref{gijeqn3}) 
gives an algebraic equation for the power $s$, which is 
easy to solve for $G \Lambda \ll 1$:
\begin{eqnarray}
\frac{(s - 1) (s - \frac23)}{s (s - \frac12)^2} 
\!\!& = &\!\!
16 \pi (G \Lambda)^2 \, f_{cr}' 
\quad \Longrightarrow
\nonumber \\
s \!\!& \approx &\!\!
\frac1{16\pi (G \Lambda)^2 \, f_{cr}'} 
\; \gg \; 1
\quad , \quad
G \Lambda \ll 1
\;\; . \label{s}
\end{eqnarray}

A more non-local scalar, which responds less rapidly to 
the slowing geometry, can be formed from derivatives of 
$X$:
\begin{equation}
\frac{S}{\Lambda} \; = \;
\frac{-g^{\mu\nu} \, 
\partial_{\mu} X[g] \; \partial_{\nu} X[g]}
{\frac{16}3 \Lambda} 
\quad\longrightarrow \quad
\frac{\dot{X}^2}{\frac{16}3 \Lambda} 
\qquad , \qquad
X[g] \, \equiv \, \frac{1}{\square} \, R
\;\; . \label{S2}
\end{equation}
This gives an end to inflation for relatively large values 
of $G \Lambda$ but still goes over to a power law with the 
asymptotic form (\ref{s}) for small $G \Lambda$. The reason 
is obvious from the form of $\dot{X}$ for $FRW$:
\begin{equation}
\dot{X} 
\quad \longrightarrow \quad
-\frac1{a^3(t)} \int_0^{t} dt' \; a^3(t') \, R(t') 
\;\; . \label{dotX}
\end{equation}
If $R$ simply vanished after some time, then $\dot{X}$ would 
decay subsequently like $a^{-3}(t)$. Hence the lag to changes 
in $R$ is of the order of a few Hubble times and inflation 
will only be ended for unrealistically large values of 
$G \Lambda$ such that the critical time $t_{cr}$ falls within 
this period. 

We also tried scalars formed by using inverses of other 
operators like the conformal d'Alembertian:
\begin{equation}
\square_c \; \equiv \; 
\square - \frac16 R 
\quad \longrightarrow \quad
-\frac1{a^2} \, \frac{d}{dt} \, a \,
\frac{d}{dt} \, a 
\;\; . \label{confsquare}
\end{equation}
and the Paneitz operator \cite{SD1,SD2}:
\begin{equation}
D_P \; \equiv \; 
\square^2 + 2 D_{\mu} 
\Bigl( R^{\mu\nu} - \frac13 g^{\mu\nu} R \Bigr) D_{\nu} 
\quad \longrightarrow \quad
\frac1{a^3} \, \frac{d}{dt} \, a \,
\frac{d}{dt} \, a \, \frac{d}{dt} \, a \frac{d}{dt} 
\;\; . \label{Paneitz}
\end{equation}
However, the result was always power law expansion like 
(\ref{s}) for $G \Lambda \ll 1$.

To avoid the magnitude problem and still end inflation 
requires that we evaluate the dynamical scalar far back 
in the past. One way of achieving this is to use the 
integral curves $\chi_{\mu}[g](x)$ of a timelike 
4-velocity field $V^{\mu}[g](x)$. We can construct such 
a 4-velocity field by taking the gradient of the 
invariant 4-volume ${\cal V}[g](x)$ of the past light-cone 
from the point $x^{\mu} = (t, {\vec x})$ back to our 
initial value surface at $x'^{\mu} = (t, {\vec x}')$ \cite{SPRPW}:
\begin{equation}
{\cal V} \; \equiv \;
\int_0^t dt' \int d^3x' \;
\sqrt{-g(x')} \,\,
\theta \Big( \! -\sigma(x;x') \Big)
\;\; . \label{Vplc}
\end{equation}
Here $\sigma(x;x')$ is $\frac12$ times the square of the 
geodesic length from $x^{\mu}$ to $x'^{\mu}$ \cite{DeWitt1}.
Because the volume of the past light-cone grows as one 
evolves into the future, its gradient is guaranteed to 
be timelike in any geometry. We define $V^{\mu}[g](x)$
as the normalized gradient of ${\cal V}[g](x)$:
\begin{equation}
V^{\mu}[g](x) \; \equiv \;
\frac{-g^{\mu\nu} \; \partial_{\nu} {\cal V}}
{\sqrt{-g^{\alpha\beta} \,
\partial_{\alpha} {\cal V} \;
\partial_{\beta} {\cal V}}} 
\;\; . \label{V}
\end{equation}
Now construct the integral curves $\chi^{\mu}[g](\tau,x)$ 
-- as a functional of the metric and an ordinary function 
of the parameter $\tau$ and a coordinate point $x^{\mu}$ 
-- so that they obey the conditions:
\begin{equation}
\frac{\partial \chi^{\mu}}{\partial \tau} 
\; = \; -V^{\mu}(\chi) 
\qquad {\rm and} \qquad 
\chi^{\mu}(0,x) \; = \; x^{\mu} 
\;\; . \label{chi}
\end{equation}
Our physical requirement on the scalar $S$ is to evaluate
it at {\it any} time $t^*$ such that inflation is still
dominant. Suppose $\tau^*[g](x)$ corresponds to this time 
as we follow the integral curves from $x^{\mu}$ back to 
the initial value surface. Then, we might define the scalar 
$S$ at point $x^{\mu}$ in expression (\ref{changeX}) to 
equal:
\begin{equation}
S(x) \; = \; \frac14 R[\chi^{\mu}(\tau^*;x)] 
\;\; . \label{S}
\end{equation}

The modification we have outlined would not change the 
initial value problem, is defined for any geometry, and 
seems to be invariant. It also solves the magnitude 
problem while preserving the general features of the 
way the original ansatz (\ref{p}) ends inflation. For 
an $FRW$ geometry and for the value of $\tau^*[g](x)$ 
associated with, e.g. $\frac{9}{10}$ of the time from
$x^{\mu}$ back to the initial value surface, the 
functional $Y[g](x)$ becomes:
\begin{equation}
Y[g](x) 
\quad \longrightarrow \quad
-\int_0^t dt' \; \frac1{a^3(t')} 
\int_0^{t'} dt''\; a^3(t'') \; R(t'') 
\times \frac{R({\scriptstyle \frac{1}{10}} t'')}{4 \Lambda} 
\;\; . \label{Ytau}
\end{equation}
Figure 1 shows the Hubble parameter versus time for this 
model with $G \Lambda = \frac1{100}$. Any of the less 
non-local models we tried evolved to the power law solution 
(\ref{s}) for this small a value of $G \Lambda$. The duration 
of the plot corresponds to 350 initial Hubble times. One can 
see that the period of oscillation lengthens towards the end 
because $\frac{1}{4\Lambda} R({\scriptstyle \frac{1}{10}} t)$ 
decreases, but $H(t)$ still oscillates with decreasing 
amplitude and the oscillations still drop below zero. 
For the enormously smaller values of $G \Lambda$ relevant 
to primordial inflation $\frac{1}{4\Lambda} 
R({\scriptstyle \frac{1}{10}} t)$ would be effectively 
constant during the oscillatory phase, so the period of 
oscillation would also be constant, just as in the simple 
model (\ref{p}). 

To recapitulate, we solve the magnitude problem by changing 
the constant $\Lambda$ in $\, G\Lambda \, \square^{-1} R \,$ 
into a factor of the dynamical quantity $R$, standing to 
the right of the inverse d`Alembertian. When the eventual
onset of matter domination perturbs our induced stress-energy
off its radiation dominated fixed point, the change will be
small -- rather than overwhelming -- because the multiplicative
factor of $\Lambda$ has been replaced by the vastly smaller
value of $R$ at nearly the time of matter domination. However, 
evaluating this additional factor of $R$ at the same time 
as the original factor of $R$ makes the source ``turn off'' 
too quickly to end primordial inflation. Our solution is to 
instead evaluate the additional factor of $R$ at a time 
$\tau^*[g](x)$ which is early enough that the Ricci scalar 
of primordial inflation is still nearly $4 \Lambda$. Although 
we set $\tau^*$ to be one tenth of the current time for the 
sake of definiteness, the only two requirements on it are: \\
{\it (i)} Near the end of primordial inflation the time
$\tau^*$ must be early enough so that we still have
$R(\tau^*) \approx 4 \Lambda$. \\
{\it (ii)} After the onset of matter domination $\tau^*$ 
must be late enough so that $R(\tau^*)$ is comparable to 
the square of the Hubble parameter at that time. \\
There are an enormous number of functionals $\tau[g](x)$ 
with these two properties and any of them would suffice. \\

${\bullet \; \;}$ {\bf Changing the Sign:}
We turn now to the sign problem. It arises because 
the Ricci scalar (\ref{Reps}) is positive during 
both inflation $(\varepsilon = 0)$ and matter 
domination $(\varepsilon = \frac32)$:
\begin{eqnarray}
q \; = \; -1 
\qquad \Longrightarrow \qquad 
R \!\!& = &\!\! +12 H^2 
\;\; , \label{Rinfl}\\
q \; = \; +\frac12 
\qquad \Longrightarrow \qquad 
R \!\!& = &\!\! +3 H^2 
\;\; . \label{Rmat2}
\end{eqnarray}
What we need is a source which changes sign from 
inflation to matter domination, and is still zero 
(or very small) during radiation domination. There 
are again many possibilities but a simple one that 
works is to change the source from the form motivated 
by (\ref{S}):
\begin{equation}
Y[g](x) \; = \;
\frac{1}{\square} \left[ \, 
R(x) \times \frac{R(\chi(\tau^*,x))}{4 \Lambda} 
\, \right]
\;\; , \label{oldY}
\end{equation}
to the following form:
\begin{equation}
Y_{\alpha}[g](x) \; = \;
\frac1{4 \Lambda} \, 
\frac{1}{\square} \left[ \,
R(x) \, R\Bigl(\chi(\tau^*,x)\Bigr) 
\, + \, 
\alpha \, \square R(x) \, \right] 
\;\; . \label{alphaY}
\end{equation}
The term proportional to $\alpha$ vanishes for 
both de Sitter inflation $(\varepsilon = 0)$ and 
pure radiation domination $(\varepsilon = 2)$, so 
it should make little difference until the onset 
of matter domination. For $t \gg t_m$ and for the
choice of $\tau^*[g](x)$ already discussed, the 
various dynamical scalars in (\ref{alphaY}) give
$(\varepsilon = \frac32)$:
\footnote{Because during matter domination 
$H(t) = \frac{2}{3t}$ and $R(t) = 3 H^2(t)$, 
it trivially follows that: $R(\frac{1}{10} t) 
= 10^2 R(t)$.}
\begin{eqnarray}
t \gg t_m 
\quad \Longrightarrow \qquad
R(t) \!\!& = &\!\!
+3 H^2(t) 
\quad , \quad 
R({\scriptstyle \frac{1}{10}} t) \; = \; 
+100 \times 3 H^2(t)
\;\; , \qquad \label{RRmat} \\
\square R(t) \!\!& = &\!\!
-\frac{27}2 H^4(t) 
\;\; . \label{boxRmat}
\end{eqnarray}
where we also used (\ref{boxReps}). There exists 
a reasonable range of $\alpha$ for which the sign 
gets reversed. \\

${\bullet \; \;}$ {\bf An Even Simpler Ansatz:}
Although expression (\ref{alphaY}) resolves the
sign problem, it introduces a new parameter $\alpha$
to the theory and, more importantly, it adds initial
value data since the second term $\alpha \square R$ 
contains fourth order derivatives of the metric.
\footnote{Unlike the first term $\frac14 R^2$ which
has the square of second order derivatives acting
on the metric.}

Both issues can be avoided by adopting the following
geometrically motivated form for the source:
\begin{eqnarray}
Y_S[g](x) \; = \;
\frac1{\Lambda} \; 
\frac{1}{\square} \; \Bigg\{ 
R(x) 
&\!\! \times &
\nonumber \\
& \mbox{} &
\hspace{-1.7cm}
\left[ \, - \, 
V^{\mu}\Bigl(\chi(\tau^*,x)\Bigr) \;
V^{\nu}\Bigl(\chi(\tau^*,x)\Bigr) \;
R_{\mu\nu}\Bigl(\chi(\tau^*,x)\Bigr) 
\, \right] \Bigg\}
\;\; , \qquad \label{newY}
\end{eqnarray}
which -- for an $FRW$ geometry and for the value of 
$\tau^*[g](x)$ associated with, e.g. $\frac{9}{10}$ 
of the time from $x^{\mu}$ back to the initial value 
surface -- becomes:
\begin{equation}
Y_S[g](x) 
\quad \longrightarrow \quad
\int_0^t dt' \; \frac1{a^3(t')} 
\int_0^{t'} dt''\; a^3(t'') \; R(t'') 
\times \frac{R_{00}({\scriptstyle \frac{1}{10}} t'')}{\Lambda} 
\;\; , \label{newYtau}
\end{equation}
where we have used that for these particular spacetimes:
\begin{equation}
FRW
\quad \Longrightarrow \qquad
V^{\mu}[g](x) \; \equiv \;
\frac{-g^{\mu\nu} \; \partial_{\nu} {\cal V}}
{\sqrt{-g^{\alpha\beta} \,
\partial_{\alpha} {\cal V} \;
\partial_{\beta} {\cal V}}}
\; = \;
\delta^{\mu}_{~ 0}
\;\; . \label{Vfrw}
\end{equation}

\vspace{0.4cm}

${\bullet \; \;}$ {\bf Late Time Acceleration:} 
The ansatz (\ref{newY}) and its $FRW$ form (\ref{newYtau})
has essentially the same time evolution with that
of the simple model (\ref{p}) during inflation, 
vanishes during radiation domination, and the 
transition to matter domination can trigger a slow, 
very small decrease in $p[g]$. This, in turn, can 
uncover a tiny portion of the bare cosmological 
constant and eventually cause a phase of acceleration.

To see this behaviour quantitatively, note that:
\begin{eqnarray}
R(t) \times R_{00}({\scriptstyle \frac{1}{10}} t) 
& = &
0 \times 300 H^2
\; = \;
0
\qquad\qquad\qquad\!\! , \quad 
10 \, t_r < t < t_m 
\label{RRt1} \\
& = &
3H^2 \times 300 H^2
\; = \;
900 H^4
\qquad , \quad 
t_m < t < 10 \, t_m 
\label{RRt2} \\
& = &
3H^2 \times 150 H^2
\; = \;
450 H^4
\qquad , \quad 
10 \, t_m < t < t_{\rm now} 
\qquad \label{RRt3} 
\end{eqnarray}
where we have used (\ref{Reps}-\ref{R00eps}). 

In view of (\ref{RRt2}-\ref{RRt3}), to compute 
the source $Y[g](x)$ given by (\ref{newY}):
\begin{equation}
Y_S[g](t) \; = \;
- \frac1{\Lambda} \,
\frac{1}{\square} \left[ \,
R(t) \, R_{00}({\scriptstyle \frac{1}{10}} t) 
\, \right]
\; \equiv \;
X_{cr} + \Delta Y_S
\;\; , \label{newY2}
\end{equation}
we must evaluate the action of the inverse 
differential operator $\square^{-1}$ on $H^4(t)$;
we might as well do so for any constant
$\varepsilon$ and for a generic lower limit of 
integration which we shall denote by $T$:
\begin{eqnarray}
\frac{1}{\square} \, H^4
\!\!& = &\!\!
- \int_T^t dt' \; \frac{1}{a^3(t')}
\int_T^{t'} dt'' \; a^3(t'') \; H^4(t'')
\nonumber \\
\!\!& = &\!\!
- \frac{1}{3 \, (1 - \varepsilon)}
\int_T^t dt' \; \left \{ \,
H^3(t') - \frac{a^3(T) \, H^3(T)}{a^3(t')}
\, \right\}
\nonumber \\
\!\!& = &\!\!
- \frac{1}{3 \, (1 - \varepsilon)}
\, \Bigg\{ 
- \frac{1}{2 \, \varepsilon} \, H^2(t) 
+ \frac{1}{2 \, \varepsilon} \, H^2(T) 
+ \frac{1}{3 - \varepsilon} \, 
\frac{a^3(T) \, H^3(T)}{a^3(t) \, H(t)}
\qquad \nonumber \\
& \mbox{} &
\hspace{5.5cm}
- \frac{1}{3 - \epsilon} \, H^2(T)
\, \Bigg\} 
\;\; . \label{boxHeps}
\end{eqnarray}
This is the full answer for any $\varepsilon$.
Of the four terms present in (\ref{boxHeps}),
the first and third are sub-dominant at late
times for matter domination $(\varepsilon = 
\frac32)$ because of their $t^{-2}$ and $t^{-1}$
behaviour, respectively. The remaining two
terms in (\ref{boxHeps}) are constant and give:
\begin{equation}
t \gg t_m 
\quad \Longrightarrow \qquad
\frac{1}{\square} \, H^4
\; = \; 
- \frac29 \, H^2(T)
\;\; . \label{boxHmat}
\end{equation}
Taking into account (\ref{RRt3}) and setting the 
lower limit of integration equal to $T = t_m$, the 
relevant part of the source (\ref{newY2}) becomes:
\begin{equation}
t \gg t_m 
\quad \Longrightarrow \qquad
\Delta Y_S[g](t) \; = \;
\frac1{ \Lambda} \, 
\left[ \, 200 \, H^2_m \, \right]
\;\; , \label{newYmat}
\end{equation}
where $H_m \equiv H(t_m)$. 

In terms of the new source $Y_S[g](t)$, the total
pressure (\ref{ptot1}) becomes:
\begin{eqnarray}
t \gg t_m
\quad \Longrightarrow \quad
p_{\rm tot} \!\!& = &\!\!
- \, \frac{\Lambda}{8\pi G} \, 
\Big\{ 1 - 8 \pi G \Lambda \;
f[-G \Lambda \, Y_S] \Big\} 
\label{newptot1} \\
& = &\!\!
- \, \frac{\Lambda}{8\pi G} \, \Biggl\{
1 - 8 \pi G \Lambda \;
f\Big[ -G \Lambda \, (X_{cr} + \Delta Y_S) \, \Big] \Biggr\} 
\qquad \label{newptot2} \\
& \simeq &\!\!
- \, G \Lambda^3 \, f_{cr}' \; \Delta Y_S 
\label{newptot3} \\
& \simeq &\!\!
- \, 200 \, G \Lambda^2 \, f_{cr}' \; H^2_m 
\;\; , \label{newptot4}
\end{eqnarray}
where in the last step we used (\ref{newYmat}).
The ratio of the predicted total pressure
(\ref{newptot4}) to the current total pressure
(\ref{pnow}) is:
\begin{eqnarray}
t \gg t_m
\quad \Longrightarrow \qquad
\frac{p_{\rm tot}}{p_{\rm now}}
\!\!& \simeq &\!\!
\frac{200}{3} \,
8 \pi (G \Lambda)^2 \times f_{cr}' \times 
\left( \frac{H_m}{H_{\rm now}} \right)^2
\qquad \label{newpratio1} \\
\!\!& \simeq &\!\!
\frac{200}{3} \,
8 \pi (G \Lambda)^2 \times f_{cr}' \times 
10^{10}
\;\; . \label{newpratio2}
\end{eqnarray}
For the exponential model analyzed in \cite{NctRpw2},
we have:
\begin{equation}
f(x) \; = \; e^x - 1
\quad \Longrightarrow \qquad
f_{cr}' \; = \;
\frac{1}{8 \pi G \Lambda} 
\;\; , \label{fcrexp}
\end{equation}
so that the pressure ratio (\ref{newpratio2}) takes
the form:
\begin{equation}
t \gg t_m
\quad \Longrightarrow \qquad
\frac{p_{\rm tot}}{p_{\rm now}}
\; \simeq \;
\frac23 \times 10^{12} \times G \Lambda
\;\; . \label{newpratioexp}
\end{equation}
Physical values of the dimensionless coupling constant 
$G \Lambda = M^4 M^{-4}_{Pl}$ easily balance the factor 
$10^{12}$ so that we can achieve equality of the predicted 
total pressure $p_{\rm tot}$ to the observed total pressure 
and, hence, account for the current acceleration of the 
universe. 

To recapitulate, the onset of matter domination causes 
our improved ansatz for the quantum stress-energy to 
move slightly off the point at which it exactly cancels 
the bare cosmological constant. We have shown that the 
shift approaches a small, positive cosmological constant 
which could easily have the right magnitude to induce 
the current phase of cosmological acceleration. For 
most of the matter dominated phase this solution is 
self-consistent because the residual cosmological constant 
is too small to alter the expansion history until very
late. Furthermore, the eventual onset of acceleration 
induces a new change in $Y_S[g](t)$, this time in the 
sense of diminishing the residual cosmological constant. 
However, the time scale for that to occur is the current 
Hubble scale, so one would not expect much effect as of yet. \\

${\bullet \; \;}$ {\bf Epilogue:} 
We have presented an improved {\it ansatz} for the 
most cosmologically significant part of the effective 
field equations of quantum gravity with a positive 
cosmological constant. This improvement of the simple 
{\it ansatz} analyzed in \cite{NctRpw2} and reviewed 
in the introduction herein, allows us to expect a 
reasonable post-inflationary time evolution into the 
present.

Prominent among the list of topics for future work is 
perturbations \cite{NctRpw3}. We need to show that the 
dynamical scalar mode of our model dumps the energy of 
oscillations into matter to reheat the universe in a 
natural way. If this happens, the improved model presented 
here can make the quantum gravity sector go quiescent 
during a long epoch of conventional radiation domination. 
The subsequent transition to matter domination might 
even give rise to something like the current phase of 
acceleration without severe fine tuning.

Furthermore, we must derive and solve the equation 
for scalar perturbations, at least enough to compute 
the scalar power spectrum. In contradistinction, the 
equation for tensor perturbations remains unchanged 
and we need only use the expansion history $a(t)$ 
predicted by our model in order to compute the tensor 
power spectrum. Finally, one also needs that there be 
no long-range scalar force at late times.

\newpage

\centerline{\bf Acknowledgements}
This work was partially supported by the European 
Union grant FP-7-REGPOT-2008-1-CreteHEPCosmo-228644, 
by NSF grants PHY-0653085 and PHY-0855021, and by the 
Institute for Fundamental Theory at the University of 
Florida.

\begin{figure}
\centerline{\epsfig{file=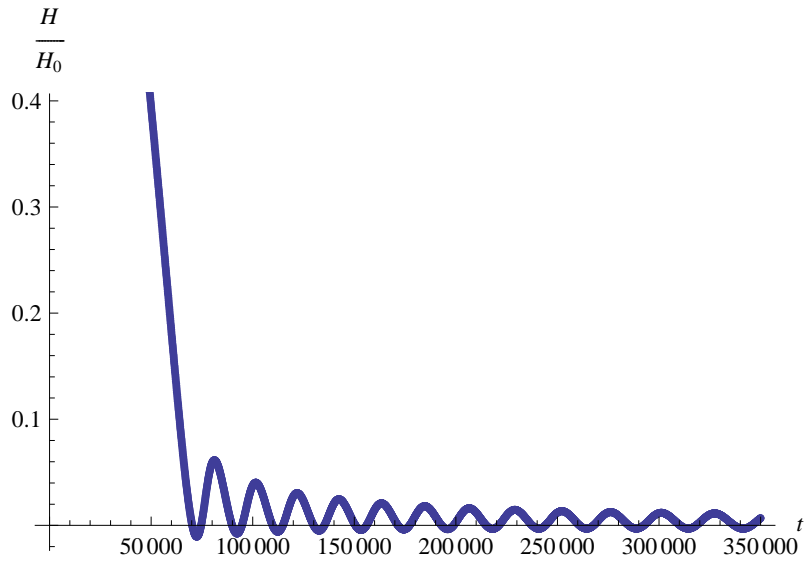,height=2.9in}}
\caption{\footnotesize The Hubble parameter (in units 
of $H_0$) versus time (in units of $\frac{1}{1000} H_0$) 
\break \mbox{} \hspace{1.95cm}
for $p = \Lambda^2 f[-G\Lambda \, Y]$ with $G\Lambda = 
\frac{1}{100}$ and $Y$ given in (\ref{Ytau}).}
\end{figure}

\end{document}